\catcode`\@=11

\newif\if@fewtab\@fewtabtrue
{\count255=\time\divide\count255 by 60
\xdef\hourmin{\number\count255}
\multiply\count255 by-60\advance\count255 by\time
\xdef\hourmin{\hourmin:\ifnum\count255<10 0\fi\the\count255}}
\def\ps@draft{\let\@mkboth\@gobbletwo
    \def\@oddhead{}
    \def\@oddfoot
     {\hbox to 7 cm{\tiny \versionno \hfil}\hskip -7cm\hfil\rm\thepage \hfil}
    \def\@evenhead{}\let\@evenfoot\@oddfoot}

\def\draftcite#1{\ifnum\draftcontrol=1#1\else{}\fi}
\def\@bibitem#1{\item\hskip -3cm \hbox to 2cm
{\hfil {\footnotesize\bf\draftcite{#1}}}\hskip 1cm
\if@filesw \immediate\write\@auxout
       {\string\bibcite{#1}{\the\value{\@listctr}}}\fi\ignorespaces}

\global\def\draftcontrol{0}
\catcode`\@=12

\def\yes{yes }  \read-1 to\yesno
\ifx\yesno\yes 
\font\tendl=msym10  scaled \magstep1
\font\sevendl=msym7 scaled \magstep1
\font\fivedl=msym5 scaled \magstep1
\newfam\dlfam \def\dl{\fam\dlfam\tendl} 
\textfont\dlfam=\tendl \scriptfont\dlfam=\sevendl
\scriptscriptfont\dlfam=\fivedl
\else \let\dl=\bf
\fi

\def\ifundefined#1{\expandafter\ifx\csname#1\endcsname\relax}
\makeatletter \ifundefined{new@mathgroup} {} \else
    \new@mathgroup\sffam
    \new@internalmathalphabet\mathsf\sffam{cmss}{m}{n}
    \def\psf{\fontfamily\sfdefault \fontseries\default@series
        \fontshape\default@shape\selectfont\mathsf}
\fi

\def\citen#1{\if@filesw \immediate\write \@auxout {\string\citation{#1}}\fi%
\@tempcntb\m@ne \let\@h@ld\relax \def\@citea{}%
\@for \@citeb:=#1\do {\@ifundefined {b@\@citeb}%
    {\@h@ld\@citea\@tempcntb\m@ne{\bf ?}%
    \@warning {Citation `\@citeb ' on page \thepage \space undefined}}%
    {\@tempcnta\@tempcntb \advance\@tempcnta\@ne
    \setbox\z@\hbox\bgroup\ifcat0\csname b@\@citeb \endcsname \relax
    \egroup \@tempcntb\number\csname b@\@citeb \endcsname \relax
    \else \egroup \@tempcntb\m@ne \fi \ifnum\@tempcnta=\@tempcntb
    \ifx\@h@ld\relax \edef \@h@ld{\@citea\csname b@\@citeb\endcsname}%
    \else \edef\@h@ld{\hbox{--}\penalty\@highpenalty
    \csname b@\@citeb\endcsname}\fi
    \else \@h@ld\@citea\csname b@\@citeb \endcsname \let\@h@ld\relax \fi}%
\def\@citea{,\penalty\@highpenalty\hskip.13em plus.13em minus.13em}}\@h@ld}
\def\@citex[#1]#2{\@cite{\citen{#2}}{#1}}%
\def\@cite#1#2{\leavevmode\unskip\ifnum\lastpenalty=\z@\penalty\@highpenalty\fi%
  \ [{\multiply\@highpenalty 3 #1%
  \if@tempswa,\penalty\@highpenalty\ #2\fi}]}   %
\makeatother 

\def\A             {Algebra}
\def\abar          {\bar A}
\def\ag            {A\gt g}

\def\bara          {\mbox{$\bar A$}}
\def\be            {\begin{equation}}

\def\bigwe         {\mbox{\small\raisebox{.2em}{$\bigwedge^{\!1}$}}}
\def\cala          {\mbox{${\cal A}\,$}}
\def\calak         {\mbox{${\cal A}_{(k)}$}}

\def\calc          {\mbox{$\cal C$}}
\def\calg          {\mbox{${\cal G}$}}
\def\calgk         {\mbox{${\cal G}_{(k)}$}}
\def\calgo         {\mbox{${\cal G}_0$}}

\def\call          {\mbox{${\cal L}$}}
\def\calm          {\mbox{${\cal M}$}}
\def\calmo         {\mbox{${\cal M}_0$}}
\def\calmk         {\mbox{${\cal M}_{(k)}$}}

\def\calo          {\mbox{$\cal O$}}
\def\caloa         {\mbox{${\cal O}_{\!A}$}}
\def\calp          {\mbox{$\cal P$}}
\def\cals          {\mbox{$\cal S$}}
\def\calsa         {\mbox{${\cal S}_{\!A}$}}
\def\calsab        {\mbox{${\cal S}_{\!\bar A}$}}
\def\calsap        {\mbox{${\cal S}_{\!A'}$}}
\def\calssm        {\mbox{\footnotesize$\cal S$}}
\def\calsnull      {\mbox{${\cal S}^{}_\circ$}}
\def\calu          {\mbox{$\cal U$}}
\def\calz          {\mbox{$\cal Z$}}
\def\citeall  {\cite{sinG,grib,koro,hkprs2,atjo,daVi,naRa,bavi0,mivi,asmi,%
              bavi,sinG2,zwan1,sefr,koSa,asmi2,hkprs,dezw3,vanb3,vava}}

\def\confi         {configuration}
\def\confisp       {configuration space}
\def\cs            {Chern\hy Simons }
\def\CS            {Chern\hy Simons }

\def\delo          {\mbox{$\nabla_{\!\!\bar A}^{}$}}
\def\delom         {\mbox{$\nabla_{\!\!A}^{}$}}
\def\deloms        {\mbox{$\nabla_{\!\!A}^*$}}
\def\delos         {\mbox{$\nabla_{\!\!\bar A}^*$}}

\def\ee            {\end{equation}}
\def\eE            {{\rm e}}

\newcommand\erf[1]{(\ref{#1})}
\def\F             {\Phi_{\!A}^{}}
\def\FA            {\Phi_{\!\bar A;A}^{}}
\def\FAA           {\Phi_{\!\bar A;\bar A}^{}}

\def\findim        {finite-di\-men\-si\-o\-nal}

\def\fmd           {fundamental modular domain}

\def\fourdim       {four-di\-men\-si\-o\-nal}
\def\FP            {Faddeev\hy Popov}
\def\fp            {\mbox{$\Delta_{\rm FP} $}}

\def\futnote#1     {\footnote{~#1}\ }
\def\futnot#1      {\ifnum\draftcontrol=1 \footnote{~internal: #1}\ \fi}

\def\gc            {{g_{\rm c}}}

\def\go            {{g_0^{}}}

\def\goo           {g_0}
\newcommand\gt[1]  {^{#1}}

\newcommand{\hsp}[1] {\mbox{\hspace{#1 em}}}
\def\hy            {$\mbox{-\hspace{-.66 mm}-}$}
\def\id            {{\sl id}}
\def\ii            {{\rm i}}
\def\iN            {\!\in\!}
\def\infdim        {infi\-nite-di\-men\-si\-o\-nal}

\def\ite           {\vspace{-.7em}\item[$\triangleright$]}
\def\itE           {\vspace{-.7em}\item[$\triangleright$]}

\def\jf            {J.\ Fuchs}

\def\konA          {\mbox{$\Omega_{\bar A}$}}

\def\LA            {\mbox{$\Lambda_{\bar A}$}}
\def\La            {\mbox{$\breve\Lambda$}}
\def\LaA           {\mbox{$\breve\Lambda_{\bar A}$}}
\long\def\labl#1   {\label{#1}\ee \ifnum\draftcontrol=1
                    \mbox{ }\\[-12 mm]\query{#1}\\[5 mm] \fi}

\def\Lanull        {\mbox{$\breve\Lambda^{}_\circ$}}
\def\Lanullb       {\mbox{$\breve\Lambda^{}_\circ\Restr{\rm bound.\,id.}$}}

\def\Largerestr    {\,\raisebox{-.8em}{\rule{.05em}{2.1em}}\,}

\def\Lh            {\mbox{$\tilde\Lambda$}}

\def\lie           {Lie algebra}

\def\ltwo          {${\rm L}^2$ }

\def\one           {\mbox{\small $1\!\!$}1}
\def\onehalf       {\mbox{$\frac12$}}

\def\pad           {\mbox{${\cal P}_{\!\rm Ad}$}}
\def\padl          {\mbox{${\cal P}_{\!\rm ad}$}}

\def\PSI           {\Gamma}

\def\qft           {quantum field theory}

\long\def\query#1{\hskip 0pt{\vadjust{\everypar={}\small\vtop to 0pt{\hbox{}%
     \vskip -13pt\rlap{\hbox to 48.9pc{\hfil{\vtop{\hsize=8pc\tolerance=6000%
     \hfuzz=.5pc\rightskip=0pt plus 3em\noindent#1}}}}\vss}}}}%
\long\def\rank#1   {\mbox{rank}\,#1}
\def\reals         {{\dl R}}

\def\rep           {representation}
\def\Rep           {Representation}

\newcommand\Restr[1] {\,\raisebox{-.3em}{\rule{.05em}{1em}}\,
                   \raisebox{-.36em}{$\scriptstyle #1$}}
\def\rmg           {e}

\newcommand\sect[1] {\section{#1}}

\def\st            {space-time}

\def\tg            {{\tilde g}}
\def\tgx           {{\gamma_x^{}}}
\def\tgxm          {{\gamma_x^{-1}}}
\def\thrpdim       {3+1\,-di\-men\-si\-o\-nal}
\def\timesG        {\!\times_G\!}
\def\tr            {{\rm tr}\,}
\def\trafo         {transformation}

\newcommand\uline[1] {{\sl #1}}
\newcommand\version[1] {\ifnum\draftcontrol=1 \typeout{}\typeout{#1}\typeout{}
                   \vspace{-13mm}\centerline{\fbox{\tt wars -- DRAFT -- #1 --
\today}}
                   \vskip3mm \fi}

\def\wrt           {with respect to }

\def\ym            {Yang\hy Mills }
\def\YM            {Yang\hy Mills }
\def\zet           {{\dl Z}}

\def\zett          {\mbox{\small {\dl Z}}}

   \newcommand{\wb}{\,\linebreak[0]} 
   \def\wB                {$\,$\wb}
   \newcommand{\Bi}[1]    {\bibitem{#1}}
   \newcommand{\Prep}[2]  {{\sl #2}, preprint {#1}}
   \newcommand{\Preps}[2] {{\sl #2}, preprints {#1}}

   \newcommand{\BOOK}[4]  {{\em #1\/} ({#2}, {#3} {#4})}
   \newcommand{\inBO}[7]  {{\sl #7}, in:\ {\em #1}, {#2}\ ({#3}, {#4} {#5}),
p.\
                          {#6}}
   \newcommand{\J}[5]     {{\sl #5}, {#1} {#2} ({#3}) {#4}}

   \def\anip  {Ann.\wb Inst.\wB Poin\-car\'e}
   \def\anop  {Ann.\wb Phys.}
   
   \def\clqg  {Class.\wb Quant.\wb Grav.}
   
   \def\comp  {Com\-mun.\wb Math.\wb Phys.}

   \def\jgap  {J.\wb Geom.\wB and\wB Phys.}

   \def\jomp  {J.\wb Math.\wb Phys.}
   
   \def\lemp  {Lett.\wb Math.\wb Phys.}
   
   \def\mpla  {Mod.\wb Phys.\wb Lett.\ A}
   
   \def\nupb  {Nucl.\wb Phys.\ B}
   \def\phlb  {Phys.\wb Lett.\ B}
   \def\phrd  {Phys.\wb Rev.\ D}
   \def\phre  {Phys.\wb Rev.}

   \def\slnm  {Sprin\-ger Lect.\wB Notes in Math.}

   \def\topo  {Topology}

   \def\WI     {{Wiley Interscience}}
   \def\WS     {{World Scientific}}

   \def\NY     {{New York}}

   \def\Si     {{Singapore}}

\documentstyle[11pt]{article}
\begin{document}  \version\versionno

\begin{flushright}  {~} \\[-23 mm] {\sf hep-th/9506005} \\
{\sf NIKHEF 95-026} \\[1 mm]{\sf May 1995} \end{flushright} \vskip 2mm

\def\thefootnote{\Alph{footnote}} \setcounter{footnote}{24}
\begin{center}
{\Large\bf THE SINGULARITY STRUCTURE OF THE}
\futnote{Invited talk at the {\sl Workshop on Quantum and Classical Gauge
Theory},
Banach Center, Warsaw, May 1995.\\ \mbox{}~~~~~ To appear in the proceedings.}
\vskip 0.2cm
{\Large\bf YANG--MILLS CONFIGURATION SPACE} \end{center}
\setcounter{footnote}{12}

\vskip 9mm{}
\begin{center} {\large J\"urgen Fuchs}\ \futnote{Heisenberg fellow}
\\[.5ex]{\small NIKHEF-H,\, Kruislaan 409}
\\{\small NL -- 1098 SJ~~Amsterdam} \end{center}

\vskip 6mm
\begin{quote}{\bf Abstract}.\\
The geometric description of \ym theories and their \confisp\ \calm\ is
reviewed.
The presence of singularities in \calm\ is explained and some of their
properties
are described. The singularity structure is analyzed in detail for
structure group SU(2).\\ This review is based on \cite{fuss}.
 \end{quote}
\vskip 8mm \setcounter{footnote}{0} \def\thefootnote{\arabic{footnote}}

\sect{\YM theory and geometry}

Even more than fourty years after the seminal paper of Yang and Mills
on `isotopic gauge invariance' \cite{yami}, a complete
characterization of those \YM theories that are relevant to
particle physics is still lacking. Some uncertainties are in fact
already present at the classical level; they include e.g.\ the
specific choice of space-time manifold and the proper
treatment of `constant' gauge transformations.

I will discuss some of these subtleties in section \ref{why}. For the
moment, however, let me to stick to one specific formulation of \YM theory,
which is based on the geometric picture of gauge symmetries. In short,
from a geometric point of view \YM theory is essentially the
\uline{theory of connections on some principal fiber bundle}.
I will describe this approach in some detail, and then proceed to
investigate its consequences for the structure of the configuration space.

Let me start by recalling that a principal fiber bundle \calp\
is a fiber bundle with total space $P$ and base space $M$,
together with the \uline{structure group}
$G$, a Lie group of dimension ${\rm dim}\,G= {\rm dim}\,P-{\rm dim}\,M$,
which satisfies the following properties. There is an action
 $\mu\!:\; P\times G\to P,\ \mu(p,\gamma) \equiv p\gamma$,
of $G$ on $P$ which is smooth and free
and is transitive on the fibers $G_x\equiv\pi^{-1}(\{x\}) \cong G$, $x\iN M$,
and the representation of $G$ on the fibers is isomorphic to the
representation of $G$ on itself that is given by right multiplication.
An \uline{automorphism} of a principal bundle \calp\ consists of a
diffeomorphism $f\!:\ P\to P$ and an automorphism $\sigma\!:\ G\to G$
such that $\mu\circ(f\times\sigma) = f \circ \mu$;
these induce a diffeomorphism $f_\pi$ of the base space $M$.
An automorphism of \calp\ is said to be \uline{vertical}, or to
\uline{cover the identity} iff $\sigma=\id_G$ and $f_\pi=\id_M$,
i.e.\ iff the diagram
  \be  \begin{array}{ccc}
  P\times G & \stackrel{f\times\id}\longrightarrow & P\times G \\[.13em]
  {\scriptstyle\mu}\,\downarrow\, \ \ && \ \ \,\downarrow \,{\scriptstyle\mu}
  \\[.13em]  P & \stackrel{f}\longrightarrow & P \\[.13em]
  {\scriptstyle\pi}\,\downarrow\, \ \ && \ \ \,\downarrow \,{\scriptstyle\pi}
  \\[.13em]  M & \stackrel{\id}\longrightarrow & M \end{array} \labl{diag1}
commutes. The vertical automorphisms of \calp\ form a group,
called the \uline{gauge group} $\calg\equiv\calg_{\cal P}$ of \calp.

An equivalent definition of \calg, which is tailored to the application
in \YM theories, is in terms of the \uline{adjoint bundle} \pad. \pad\
is defined as the bundle over $M$ associated to \calp\ that has
total space $P\timesG G$
(the elements of $P\timesG G$ are classes $[p,\gamma]:=\{(p\tilde\gamma,
\tilde\gamma^{-1} \gamma \tilde\gamma\,|\,\tilde\gamma\in G\}$)
with the projection defined as $\pi_{\rm Ad}^{}([p,\gamma]):=\pi(p)$.
The gauge group \calg\ is then the set of sections $s$ of \pad, with
composition law $s\cdot s'(x)=[p,\gamma\gamma']$
for $s(x)=[p,\gamma]$ and $s'(x)=[p,\gamma']$, for all $x\iN M$.
In this description, the Lie algebra \call\ of \calg\ is the space of
sections of an analogous vector bundle \padl\ with total space
$P\timesG L$, where $L$ is the \lie\ of $G$.
According to this definition the elements of \calg\ can
be interpreted locally as $G$-valued smooth functions.
If the bundle \calp\ is trivial (i.e.\ $P\cong M\times G$), then
this interpretation is valid globally. It must be stressed, however,
that even in the latter situation there does not exist any canonical embedding
of the structure group $G$ into the gauge group \calg.

A natural additional structure on a principal fiber bundle is provided
by the notion of a \uline{connection}, which allows for the definition of
parallel transport and of covariant derivatives of sections in associated
vector bundles. In physical terms, these are important ingredients of the
dynamics; first, the connection provides the basic dynamical variables,
the gauge fields, and second, the covariant derivatives allow for the
consistent coupling of matter to the gauge fields.

Geometrically, a connection $A$ consists in a particular decomposition of
the tangential space $T_pP$ of $P$ (at any point $p\iN P$) into a
direct sum $H_p\oplus V_p$ of orthogonal `horizontal' and `vertical' spaces.
Algebraically, $A$ is a differential one-form over $P$ with
values in $L$ which under the action of $G$ on $P$ transforms according to the
adjoint \rep\ of $G$, such that the horizontal space $H_p$ is the kernel of
$A_p$. The most convenient characterization of $A$ for the application to \YM
theory is as a set $\{A_{\cal U}^{}\}$ of $L$-valued one-forms over \calu,
indexed
by an atlas $\{\calu\}$ of $P$ with appropriate transition functions.
These give rise locally to one-forms over $\calu_M$, with $\{\calu_M\}$ an
atlas of
the base space $M$, and hence in the case of trivial bundles to a
one-form over $M$.
The space of all connections on the bundle \calp\ will be denoted by \cala.
The space \cala\ (together with an action of the vector space
$\bigwe\!\otimes L$ of equivariant one-forms over $P$ with values in $L$)
is an affine space; after the choice of a base point \bara, \cala\ can be
interpreted as a (real, \infdim) vector space according to
$\cala=\{\bara+a\,|\,a\in\bigwe\!\otimes L\}$.

Any gauge \trafo\ $g\iN\calg$ can be pulled back from \calp\ to \cala.
The pull-back $g_*$ acts locally as
  \be  g_*(A)=A^g := g^{-1}Ag+g^{-1}{\rm d}g \,.\labl{g*}
On the vector space associated to \cala, \calg\
acts homogeneously, $\bigwe\!\otimes L\ni a\mapsto g^{-1}ag$.

\sect{The \confisp\ \calm}

Having introduced the geometric concepts above, I can now
present the definition of \ym theory, or more precisely, of its
`kinematical part'. This is described by the
space \cala\ of connections on a principal fiber bundle \calp\ and the action
\erf{g*} of the gauge group \calg\ on \cala, where the structure group $G$
is required to be a simply connected (semi-)simple compact matrix
Lie group, and the base space $M$ to be a `space-time'. The latter term
just refers to the requirement
that $M$ should be of potential interest to applications in physics, but
otherwise is by no means precise. In the following I take $M=S^4$, a choice
to be justified later on.
The \uline{configuration space} of the \YM theory is the space
  \be  \calm := \cala / \calg   \ee
of connections modulo gauge transformations,
or more explicitly, the \uline{orbit space} $\calm = \{ \caloa \mid
A\in\cala\}$,
whose points are the \uline{gauge orbits}
  $  \caloa := \{ B \iN \cala \,|\, B=A^g \ \mbox{for some}\ g \iN \calg \}$.

For $M=S^4$, the isomorphism classes of principal
$G$ bundles \calp\ are labelled by an integer $k$,
the instanton number or Pontryagin class of \calp.
 \futnote{Similarly, for $M=S^2\times\mbox{\small\reals}$ and structure group
U(1), the isomorphism classes of principal bundles are again labelled by an
integer, the monopole number.}
 For the geometric description of \ym theory one must choose
 \futnote{There may actually be subtleties to this choice, as the family
of isomorphic bundles need not be a set.}
a representative of a
 fixed isomorphism class, and hence fix the instanton number to
a definite value $k$. Actually, for many of the aspects I will mention below,
the relevant value will be $k=0$, so that \calp\ is a trivial bundle.
It should be noted
that for any fixed $k$ there is a separate space \calak\ of
connections, each being acted on by a separate gauge group \calgk,
and hence a separate \confisp\ $ \calmk = \calak/\calgk$.
For some applications in physics it is actually necessary to
allow for arbitrary $k\in\zett$; doing so, one deals with
a `total configuration space' which is the disjoint union over all \calmk,
  \be  \calm_{\rm tot} = \bigcup_{k\in\zet}^. \calak/\calgk \,. \ee
{}From now on, I will always refer to a definite bundle at fixed
instanton number, and correspondingly will suppress the label $k$.

Let me remark that
the \uline{\infdim} geometry of the spaces \cala\ and \calm\ is
conceptually rather different from the finite-di\-mensional geometry of a
specific gauge field configuration, i.e.\ of a single fixed connection
(it is often the latter which is referred to as `the geometry of \YM theory').
However, under favourable circumstances the local geometry
of \calm\ is closely related to global properties of the base manifold.
For instance, if one restricts to the \uline{\findim}
solution set of suitable differential
equations, for $G={\rm SU}(2)$ and $M$ a compact four-manifold there are
the relations described by Donaldson theory \cite{dona4,DOkr}
and Seiberg\hy Witten theory \cite{witt45,eicH}.

To complete the definition on \ym theory, I still have to prescribe its
`dynamical part'. This is done by specifying an action functional
$S_{{\rm YM}}\!\!:\;\cala\to\reals$. For definiteness, this will be taken to
be the ordinary \YM action
  \be S_{{\rm YM}} [A] = \frac1{4{\rmg}^2}\, |F|^2 \,,\ee
where g is a coupling constant, $F={\rm d}A+\onehalf\,[A,A]$\, is the curvature
of the connection $A$, and
$|\cdot|$ denotes the \ltwo norm on equivariant $p$-forms on
$M$ with values in $L$ that is induced by the \ltwo scalar product
  \be   (B,C) := \int_M {\rm tr} (B \wedge *C)\, . \labl{norm}
(Thus in local coordinates, $4{\rmg}^2\,S_{{\rm YM}} [A] =
\int_M {\rm d}^4x\, {\rm tr} (F_{\mu\nu} F^{\mu\nu})$
with $F_{\mu\nu}(x) = \partial_\mu A_\nu(x) - \partial_\nu A_\mu(x)
+[A_\mu(x),A_\nu(x)]$ and $A=\sum_{\mu=1}^4A_\mu(x){\rm d}x^\mu$.)

In the analysis below the specific form of the action functional will
not play any particular r\^ole. Rather, the only really relevant property
of $S_{{\rm YM}}$ is that it is gauge invariant in the sense that
  \be  S_{{\rm YM}} [A\gt g] = S_{{\rm YM}} [A] \labl{g.i}
for all $g\iN\calg$.
For most of the considerations below, one could therefore also have in
mind any other action functional sharing this crucial property, such as
in three dimensions the \cs action or \cite{afll2} the combination of
\cs and \ym actions.
The gauge invariance \erf{g.i} implies that the action is in fact a
well-defined functional $S_{{\rm YM}}\!:\; \calm\to\reals$
on the \confisp\ \calm\ rather than just on the `pre-\confisp' \cala.

\sect{Gauge fixing}

While \cala\ is an affine space and hence
easy to handle (e.g.\ it is contractible), the structure of the \confisp\
\calm\ is much more complicated. For instance, \calm\ is not a manifold
(see below), and even when restricting to manifold points, at least
one homotopy group is non-trivial.
It is therefore most desirable to describe \calm\ as concretely
as possible in terms of the simple space \cala.
(Often it is in fact advantageous to allow in intermediate steps for
quantities which are not well-defined on \calm, e.g.\ this can help
to formulate the theory in a `less non-linear' way.
This option is in fact one of the basic reasons for dealing with
gauge theories in the redundant description in terms of connections.
In the present context, one should avoid the use of
gauge-noninvariant objects whenever possible.)

One of the key ideas in approaching the space \calm\ via \cala\ is to
identify a subset of \cala\ that is isomorphic to \calm\ modulo boundary
identifications. Such a subset is called a \uline{\fmd} and will
generically be denoted by $\Lambda$. A prescription for
determining a \fmd\ $\Lambda$ (referred to as \uline{fixing a gauge})
consists in picking in a continuous manner
a representative out of each gauge orbit $\caloa$. Technically, one tries
to achieve this by considering the set $\{ A \in \cala \mid \calc[A]=0 \}$
for a suitable functional \calc\ on \cala. This set is called a
\uline{gauge slice} associated to the gauge condition \calc\ and will be
denoted by $\Gamma$. In a geometrical context the natural gauge condition is
the \uline{background gauge}
  \be  \calc[A]\equiv \calc_{\!\abar}[A] := \nabla_{\!\!\abar}^*(A-\bara) \,,
  \labl{calc}
where $\bara$, called the background connection,
is an arbitrary element of \cala. (In the special case of $k=0$ and
background $\bara=0$ in which \erf{calc} reduces to $\calc[A]={\rm d}^*\!A$,
this is also known as the Lorentz, or Landau gauge, or \cite{uhle} as
the Hodge gauge.) Thus the gauge slice is the subset
  \be \Gamma \equiv \Gamma_{\!\abar} = \{ A \in \cala \,|\,
  \nabla_{\!\!\abar}^*(A-\bara)=0 \}  \ee
of \cala. As already indicated by the use of a different symbol $\Gamma$ in
place of $\Lambda$, \erf{calc} does in fact not provide a complete
gauge fixing. This failure is not an artifact of the background gauge, but
(at least for $M=S^4$, and presumably for any compact $M$)
happens for any continuous gauge condition \calc\ \cite{sinG}.
(There do exist continuous gauge conditions which avoid this problem
on a large subset of \cala, though not globally on \cala,
such as the axial gauge on $M=\reals^4$ \cite{chod}
and axial-like gauges on tori \cite{lask}.)

Thus  for any $\abar\iN\cala$ there exist
orbits \calo\ containing distinct connections $A$ and $B$ such that both
$A$ and $B$ lie in $\Gamma_{\!\abar}$;
$A$ and $B$ are then called \uline{gauge copies} or \uline{Gribov copies}
\cite{grib} of each other. Gauge copies appear at least outside a subset
$\Omega\equiv \konA$ of $\Gamma$ (that is,
any $A\in\Gamma \setminus \Omega$ has a gauge copy within $\Omega$) which
can be described as the set of those connections for which $g=e$
(the unit element of \calg) is a minimum of the functional
  \be  \F \equiv \FA[g] := |A^g -\bara|^2 \ee
on \calg. The functionals $\F$ also contain the (unphysical)
information about the topology of the orbit \caloa\ and correspondingly
are sometimes referred to as Morse functionals.
The subset $\Omega$ is known as the \uline{Gribov region}.
Any orbit \calo\ intersects
$\Omega$ at least once, and $\Omega$ is convex and bounded.
The r\^ole of the Morse functionals $\F$ is best understood by observing
that
  \be  \hsp{-1}\frac{\delta \F}{\delta g}\Largerestr\raisebox{-.8em}
  {$\scriptstyle g=e$}\!\! = -2\,\deloms(A-\bar A) = -2\,\delos(A-\bar A)
  \,,  \qquad
  \frac{\delta^2 \F}{\delta g^2}\Largerestr\raisebox{-.8em}{$\scriptstyle
  g=e$}\!\! = 2\,(\delo w,\delom w) = - 2\,(w, \delos\delom w) \,. \labl+
Thus the vanishing of the first variation of $\F$
yields the gauge condition, so that
for all $A\in \Gamma$, $e\iN\calg$ is a stationary point of $\F$, while the
Hessian of the variation is given by the \FP\ operator $\fp= - \delos \delom$,
implying that $A\iN\Omega$ for all $A\iN\Gamma$ for which this operator is
positive.

The Gribov region is not yet the modular domain $\Lambda$, as is
indicated by the choice of the different symbol $\Omega$. However
for \uline{generic} background $\bara$ it already comes
rather close to $\Lambda$. Namely,
in the definition of $\Omega$ both absolute and relative minima of the Morse
functionals $\F$ contribute, and $\Lambda$ is obtained
by just restricting to the absolute minima,
  \be  \Lambda \equiv \LA=\{A\in \PSI \mid \F[g]\geq\F[e] \
  \mbox{for all}\ g\in\calg\} \subseteq \konA \,. \labl L
The set \erf L contains at least one representative
of each gauge orbit, provided that one considers the gauge
group \calg\ as completed in the \ltwo norm. (It is not known whether this
remains true when completing \wrt an arbitrary Sobolev norm.)
Also, the interior of $\LA$ contains (for generic $\bara$)
at most one representative, i.e.\
gauge copies only occur on the boundary $\partial\LA$. Thus for
generic background, $\LA$ is a \fmd. Furthermore,
just as the Gribov region $\Omega$, its subset $\Lambda$ is
convex and bounded, and (for any compact \st\ with
$H^2(M,\zet)=0$) it is properly contained in $\Omega$.

The action of the gauge group \calg\ on \cala\ is not free, i.e.\
the \uline{stabilizer} (or isotropy subgroup)
  $ \calsa :=\{ g\in\calg \mid A^g=A\} $
of a connection $A$ may be non-trivial. Indeed,
for all $A\iN\cala$ the stabilizer contains the group \calz\
of constant gauge transformations with values in the center $Z$ of $G$.
In the statements made above, the qualification of the background $\bara$
as `generic' means that $\calsab\cong\calz$.
Connections with this property are also referred to as \uline{irreducible},
while connections $A$ for which \calsa\ contains \calz\ as a proper subset are
called \uline{reducible}.

Some properties of the stabilizer groups are the following.
Any stabilizer \calsa\ is isomorphic to the centralizer ${\cal C}(H_A(p))$
of the holonomy group of the connection $A$, and accordingly is
isomorphic to a closed Lie subgroup of the structure group $G$, and hence
in particular \findim.
Within any fixed orbit \calo, all stabilizers \calsa, $A\iN\calo$, are
conjugate subgroups of \calg, i.e.\ for any $A,B\iN\calo$ there exists an
element $g\iN\calg$ such that $\cals_B=g^{-1}\calsa g$.
For any \st\ $M$ and any compact $G$,
the set of all such conjugacy classes is countable \cite{koro}.
Finally, $\calsa\cong G$ iff \bara\ is a pure gauge, i.e.\ iff $A=g^{-1}
{\rm d}g$.

The gauge invariance of the \ltwo norm $|\cdot|$ implies that
   $ \F[gh] = |A\gt{gh} - \abar|^2 = |A\gt{gh} - \abar\gt h|^2
     = |(\ag - \abar)\gt h|^2 = \F[g] $
for any $h\iN\calsab$. Because of this systematic
degeneracy, in particular the absolute minima are degenerate,
which means that in fact not $\LA$, but rather the quotient
   \be  \LaA := \LA / \calsab \ee
is a \fmd. For irreducible \bara\ this is, however, irrelevant, since then
$\calsab=\calz$ and \calz\ acts trivially, so that $\LaA= \LA / \calz= \LA$.

The distinction between reducible and irreducible backgrounds also appears in
various other circumstances. For instance, if
\bara\ is irreducible, then the Gribov region $\Omega$ can be described as
the set $\{ A\in\PSI \mid(A,\fp A) \geq 0 \}$
in which the \FP\ operator is positive, whereas for reducible \bara,
det($-\delos\delom)$ vanishes identically (but even then $\Omega$
can be described in terms of the Morse functionals $\F$ as above).

\sect{The stratification of \calm}

For any subgroup \cals\ of \calg,
let $[\cals]=\{\cals'\subseteq\calg\mid\cals'=g^{-1}\cals g,\
g\iN\calg\}$ denote its conjugacy class in \calg.
The set of all orbits $\caloa\iN\calm$ which have
fixed stabilizer type $[\cals]$, i.e.\ whose elements $A\iN\caloa$
have stabilizers $\calsa\iN[\cals]$, is a \uline{Hilbert manifold}, i.e.\ an
infinite-dimensional $C^\infty$ manifold modelled on a Hilbert space.
However, the full \confisp\ \calm\ is not a manifold, but rather it has
singularities, which are at the orbits of reducible connections.
More precisely \cite{koro}, \calm\ is a \uline{stratified variety}. Thus
as a set \calm\ is the disjoint union of (countably many)
smooth manifolds, the strata of \calm.
Any stratum with stabilizer type $[\cals]$ is dense in the union of all strata
that have stabilizers containing some $\cals'\iN[\cals]$. In particular, the
\uline{main stratum}, i.e.\ the stratum of orbits of irreducible connections,
is dense in \calm. Also,
each stratum can be described as the main stratum of another \confisp\ that
is obtained from the space of connections on some subbundle of \calp\
\cite{hkprs2}. \vskip2mm

Let me pause to point out that almost all results I described so far can be
found, though widely scattered, in the literature. Some of the
main references are \citeall. However, a few
of the details I presented were found quite recently \cite{fuss}; the results
reported below are again based on \citeall, but
to a large extent have been obtained in \cite{fuss}.\vskip2mm

I will first describe the stabilizers of the connections within the domains
$\LA$ in more detail. (This completes the proof of the claim
that $\LaA=\LA/\calsab$ is a \fmd, or in other words, that the degeneracy of
the functionals $\F$ is already exhausted by
the systematic degeneracy associated to $\calsab$.)
Given the domain $\Lambda$, define \Lh\ as
the subset of connections in $\Lambda$ that only have the systematic
degeneracy.
The functional $\FAA[g] = |\bar A^g -\bar A|^2 = |g^{-1}\delos g|^2 $ attains
its absolute minimum, namely 0, iff $g\iN\calsab$. Thus $\bara\iN\Lh$, hence
in particular \La\ is not empty. Similarly, for any
$B\iN\Lambda\!\setminus\!\Lh$
one can show that the straight line between $B$ and \bara\ (except for the
point $B$ itself) is contained in \La, so that $\Lambda\!\setminus\!\Lh \subset
\partial\Lambda$. On the other hand, if for some $A\iN\Lambda$ the stabilizer
\calsa\ is not contained in \calsab, then there is an additional degeneracy,
so that $A\in\partial\Lambda$. In particular, if for some
$A\iN\Lambda\!\setminus\!\partial\Lambda$, $\caloa$ belongs to the
same stratum as $\calo_{\!\bar A}$, then the stabilizers are not just
conjugated, but in fact coincide, $\calsa =\calsab$. Thus the interior of
$\Lambda$ contains only connections whose stabilizer is either identical or
strictly contained in that of \bara. In particular, for irreducible \bara,
all reducible connections in $\LA$ lie on the boundary $\partial\LA$.

\sect{Boundary identification and geodesic convexity}

The \fmd\ $\LA$ is isomorphic to \calm\ only modulo boundary identifications,
which account for the topologically non-trivial features of \calm.
To describe some aspects of the required boundary identifications,
consider the modular domain $\LA$ for an irreducible background
\bara, and a point $B$ on the boundary $\partial\LA$ that is irreducible
as well. Next regard $B$ instead as an element of the modular domain
$\Lambda_B$
which is isomorphic to \calm\ modulo boundary identifications, too.
$B$ is an inner point of $\Lambda_B$ and hence corresponds to a smooth
inner point of \calm. This implies that upon boundary identification of $\LA$,
a neighbourhood of $B$ on $\partial\LA$ (consisting of
irreducible connections only, since reducible connections are nowhere dense)
gets identified with another neighbourhood on $\partial\LA$.

In contrast, if $B$ is reducible, then upon proceeding from $\Lambda_B$ to
the modular domain $\Lambda_B/\cals_B$, $B$ becomes a singular point,
and this remains true upon boundary identification. Thus when $B\iN\partial
\LA$ for irreducible \bara, then in the boundary identification process $B$
must again become a singular point; as a consequence, there are `less' boundary
identifications for reducible elements of $\partial\LA$ than for irreducible
elements. As reducible connections cannot be boundary points of
$\LA\Restr{\rm bound.\,id.}$, and hence of the \confisp\ \calm\
(the codimension of the reducible strata is infinite), it follows in
particular that \calm\ does not possess any boundary points.

As, contrary to $\LA$, the set $\LA/\calsab\Restr{\rm bound.id.}$ is no longer
a subset of an affine space, there is no notion of convexity any more.
But as each of the strata is a Hilbert manifold, there is still the notion
of \uline{geodesic convexity}, meaning that
any two non-singular points can be joined by a geodesic which only
consists of non-singular points (in singular points geodesics cannot be
defined). Now the \uline{main} stratum of \calm\ is geodesically convex.
(But it is not known whether the non-main strata are geodesically convex.)
To see this, take
$P_A, P_B \iN\LA/\calsab$ non-singular and let $B$ be a representative
of $P_B$ in \LA\ and $C$ a representative
of $P_C$ in $\Lambda_B$. As $\Lambda_B$ is convex, the straight line
from $B$ to $C$ is contained in $\Lambda_B$ and (since $B$ is irreducible)
contains no reducible connection.
This line gets projected to a geodesic in $\LA/\calsab$ \cite{bavi}.
This is still true if $C$ is an irreducible connection on the boundary
$\partial \Lambda_B$, and hence also after boundary identification.

\sect{SU(2) \ym theory}

In this section I specialize to $G={\rm SU}(2)$ (and $M=S^4$), which allows
to produce several rather concrete results. In this situation,
the instanton number of any reducible connection vanishes \cite{sinG},
so that \calm\ has singularities only for $k=0$. Correspondingly I will only
consider $\calm\equiv\calm_{k=0}$; the connections and gauge \trafo s can then
be described as smooth functions on $M$.

For structure group SU(2), there are only three possible stabilizer types,
namely either $\calsa=\calz\equiv Z(\mbox{SU}(2))=\{\pm\one\}$,
in which case $A$ is irreducible;
or $\cals\cong{\rm SU}(2)$, then $A$ is a pure gauge;
or else $\cals\cong{\rm U}(1)$. An obvious task is to classify
the conjugacy classes \wrt \calg\ which correspond to each of the three
isomorphism classes of stabilizers. It turns out that in the SU(2) case
there is a unique stratum for each of the three stabilizer types.

Each orbit of connections with stabilizers \calsa\ of type $[{\rm U}(1)]$
contains one representative $A'$ for
which \calsap\ consists of constant gauge transformations with values in U(1).
Namely, in any fiber of \calp\ one has the freedom to multiply by an element
of SU(2), and the only question is whether this can be extended globally
in the appropriate way.
Now $g\iN\calsa$ implies $\nabla_{\!\!A} g=0$. By considering $g$ close
to the identity of \calsa, it follows that $\delom\sigma=0$
for some element $\sigma$ of the \lie\ \call\ of \calg, and hence
$\ell:= \tr (\sigma(x)\sigma(x)^+)$ is constant on $M$.
SU(2) acts transitively on elements $\sigma(x)\iN {\rm su}(2)$
that have a fixed value of $\ell$, and hence
one can fix a $\tau$ in su(2) with ${\rm tr}(\tau\tau^+_{})=\ell$
such that for any $x\iN M$ there is an element $\tgx\iN{\rm SU}(2)$ with
       $  \sigma(x) = \tgx \tau \tgxm$.
Provided that setting $\tg(x):=\tgx$ for all $x$ yields a well-defined gauge
\trafo, it follows that $\tau$ lies in the \lie\ $\calssm_{\!\!A'}$ of the
stabilizer $\cals_{\!\!A'}$ of $A':=A^{\tg^{-1}}_{}$,
and hence $\cals_{\!\!A'}$ is the U(1) generated by $\tau$, which consists of
constant gauge transformations. Thus the representative whose existence I
claimed is the $\tg$-transform $A'$ of $A$.
Now for $M=S^4$ the above prescription does provide a well-defined
$\tg\iN\calg$. The only ambiguity in $\tgx$, and hence the only
potential obstruction, is in the stabilizer U(1) of $\tau$,
and as one can smoothly continue $\tg(x)$ as long as the topological
non-triviality of $M$ is irrelevant, the obstruction is in fact
parametrized by the set of maps from $S^3\subset S^4$ to $U(1)$, i.e.\ the
homotopy group $\pi_3({\rm U}(1))$, which however vanishes.
(More generally, for $M=S^n$ the obstruction is parametrized by
$\pi_{n-1}({\rm U}(1)) = \delta_{n,2}\zet$.)

To summarize, for any orbit of connections with U(1) stabilizer type there is
a representative $A'$ of the form
       \be  A'_\mu(x) = \tau a_\mu(x) \labl{above}
with ordinary functions $a_\mu$. In particular, the theory has only one
\uline{single} U(1) stratum.
For arbitrary semi-simple structure group $G$,
analogous considerations apply to the particular stratum whose
stabilizer type is the maximal torus of $G$.
(In contrast, one must expect an infinite number of strata whose stabilizer
type is a proper non-abelian Lie subgroup of $G$.)

One may also analyze which of the specific connections $A^g$ of the
form \erf{above} lie in the gauge slice. Choosing for simplicity
the background $\bar A=0$, the gauge condition then reads explicitly
$\tau\partial^\mu a_\mu = [\partial_\mu g\,g^{-1},\tau]a_\mu-\partial^\mu
(\partial_\mu g g^{-1})$, which
as a special class of solutions has $g(x) = \exp (\ii\tau\gamma(x))$, where
$\gamma(x) = -\ii\int_M {\rm d}^4y H(x,y) \partial^\mu a_\mu(y) +{\rm const}$\,
with $H(x,y)$ the Green function of the Laplacian on $M$. But the
\uline{general}
solution of the differential equation for $g$ is less trivial; in
particular one cannot easily determine which solutions
lie in the \fmd\ $\Lambda$, except
in the special case where the connection is flat,
where the only representative of the orbit of \erf{above} that is contained
in $\Lambda$ is the background $A=0$ itself. The flat case is precisely the
one in which the stabilizer type gets extended to SU(2), and $A=0$ is the
only point on that orbit for which the stabilizer consists precisely
of the constant gauge transformations.

In fact, the \fmd\ $\Lambda$ is not a particularly convenient tool for the
global description of the reducible connections.
Fortunately, in the specific case of SU(2), another characterization is
available, namely via a relation with the configuration space of
electrodynamics. As any principal U(1)-bundle over $S^4$ is trivial, one can
consider $a=0$ as a base point in the space $\cala_1$ of connections of the
U(1)-bundle, so that $\cala_1$ can be viewed as the vector space of (4-tuples
of) smooth functions $a_\mu$ on $S^4$. The gauge group then acts on these
functions as $a\mapsto a+\ii\,{\rm d}\lambda$ with $\lambda\iN
C^\infty(S^4,\reals)$. Except for constant $\lambda$ which leaves all $a$
invariant, this action is free, so that the \confisp\ of the U(1) gauge
theory is a manifold.
Now fixing an element $\tau$ in the \lie\ su(2), the map
      \be  \phi:\quad \cala_1\to\cala\,, \qquad a\mapsto A=\tau a \labl{1-2}
is well-defined on orbits because the images of $a$ and $a+\ii\,{\rm d}\lambda$
are related as $\tau(a+\ii\,{\rm d}\lambda)=A^g$ with
the SU(2) gauge \trafo\ $g=\eE^{\ii\lambda\tau}$. However, in the
U(1) theory $a$ and $-a$ lie on distinct orbits (for $a\ne0$), while in the
SU(2) theory $\tau a$ and $-\tau a$ are on the same orbit. Thus
the mapping \erf{1-2} is not one-to-one on orbits, but two-to-one, and
hence the U(1)
stratum of the \confisp\ \calm\ of SU(2) \ym theory is a $\zett_2$-orbifold
of the configuration space $\calm_1$ of the U(1) theory.
The SU(2) stratum is the unique fixed point of the $\zett_2$ action.

Further analysis shows that \calm\ has a `cone over cones' structure. The
U(1) connections form an infinite-dimensional cone with tip $A=0$, and any
orbit with U(1) stabilizer is itself the tip of an infinite-dimensional cone
whose base consists of irreducible connections.

\sect{The pointed gauge group}

The \uline{pointed gauge group}, defined by
  \be \calgo := \{ g\in\calg \mid g(x_0) = e \in G \} \ee
with $x_0\in M$ fixed, acts freely on the space \cala\ of connections.
Thus the associated orbit space $\calmo := \cala / \calgo$\,
is a manifold (and accordingly is popular among mathematicians).
For instanton number $k=0$, any $g\iN\calg$ can be written as $g =\gc\go$
with $\go\iN\calgo$ and $\gc$ the constant gauge transformation with
$\gc(x)=g(x_0)$ for all $x\iN M$. One can show that the map
  \be  \varphi:\quad \Lanullb\to  \calmo \,,\qquad
  \Lanull\ni A \mapsto [A \bmod \calgo] \in \calmo \ee
(with $A\iN\Lanullb$ considered as an element of $\Lanull:=\breve\Lambda
_{\bar A=0}$)
is a diffeomorphism between \Lanullb\ and \calmo\ and intertwines the
group action of $G$ on these spaces.
(On both spaces, $G$ acts via constant gauge \trafo s $\gc$, namely on
$\Lanull$ as $A\mapsto A^\gc$, which is well-defined because
$\gc\iN\calsnull$, and on \calmo\ as
$[A\bmod\calgo]\mapsto [A^\gc\bmod\calgo]$, which is well-defined
because \calgo\ is a normal subgroup of \calg\ so that
$B^\gc=A^{\go\gc}=A^{\gc\goo'}$ with $\goo'=\gc^{-1}\go\gc$ for $B=A^\go$.)
The intertwining property of $\varphi$ means that
$\varphi( A^\gc)= [A^\gc\bmod\calgo]$, which immediately follows from the
definition of $\varphi$.

\sect{Motivations} \label{why}

Let me come back to the issue of the relevance of
defining \YM theories in the way described above. Thus
I must explain the various choices I made, and also
discuss whether there are any effects which distinguish between these
and different choices.
Concerning the first task, I have to justify the specific choice of \st\
and the definition of the gauge group. While I cannot offer any fully
convincing
arguments, the following aspects certainly play an important r\^ole.
\\[.3em]
The reasoning leading to $M=S^4$ goes as follows. First, \st\ is taken to be
euclidean on the following grounds.
  \begin{itemize}
  \itE{  I have in mind a lagrangian framework, and ultimately a
         description of the \uline{quantum} theory in terms of path integrals,
         which have a better chance of being well-defined if $M$ is euclidean.}
  \itE{  The \fourdim\ treatment appears to depend to a smaller extent on the
         classical dynamics (i.e.\ choice of the action). In the \thrpdim\
         approach the constraints which enforce the reduction of the naive
         `pre-phase space' (spanned by three-dimensional gauge fields and
         their conjugate momenta, the electric field strengths) to a
         constraint surface are part of the \fourdim\ equations of motion,
         and hence the classical dynamics enters already at an early stage.}
   \end{itemize}
\vspace{-.6em}
Also, I feel a bit uneasy about the Schr\"odinger picture employed in
the \thrpdim\ approach to approach the quantum theory; as it is so close to
quantum mechanics, one potentially misses such significant features of
\qft\ as superselection rules.\\[.3em]
Second, some reasons for compactifying $\reals^4$
to $S^4 \cong \reals^4 \cup \{ \infty \}$, the topological one-point
compactification of $\reals^4$ (and to consider $S^4$
as endowed with its natural metric so that it has finite volume), are the
following.
  \begin{itemize}
  \itE   Requiring the action, and hence $|F^2|$, to be finite, one needs
         $F\to0$ for $|x|\to\infty$, and hence $A\to g^{-1}{\rm d}g$
         (pure gauge), i.e.\ $\caloa\to\calo_0$.
         Thus the compactification to $S^4$
         corresponds to natural boundary conditions at infinity.
  \itE   Instanton sectors (i.e.\ bundles with $k\ne0$) only appear for
         compact $M$. Even though still poorly understood, topologically
         non-trivial gauge field configurations are expected to play a
significant
         r\^ole for various nonperturbative features of \qft, ranging from the
         structure of the QCD vacuum to baryon number violation in the
         electroweak interactions.
  \itE   Having finite volume has technical advantages, e.g.\ concerning the
         normalizability of constant gauge transformations.
         (This is important whenever one needs on \calg\ not only the
         group structure, but also a topological structure, such that \calg\
         becomes an infinite-dimensional \uline{Lie} group.)
  \itE   Finite volume automatically provides an infrared regularization.
         This is often convenient, as it is to be expected that any detailed
         study of the quantum theory involves (at intermediate steps)
         the introduction of an infrared cutoff.
  \end{itemize}
\vspace{-.3em}
Finally one has to specify which transformations are to be counted as gauge
transformations in the sense of \uline{redundancy}
transformations, i.e.\ which \confi s of the basic variables (here the
connections $A$) are to be considered as equivalent and hence as
describing the same physical state.
Thus in particular I have to argue why to take the full gauge group \calg\
(rather than
e.g.\ the pointed gauge group \calgo), and why to take the full \confisp\
(rather than e.g.\ restricting to the main stratum).
An obvious reason for this choice is that it is just
the most straightforward thing to do. Other arguments include:
  \begin{itemize}
\itE  \calgo\ depends on the choice of base point $x_0$. Even though one
  stays within a definite isomorphism class, in principle the choice of \calgo\
  might introduce an unwanted dependence of the physics on the \st\
  point $x_0$. (In physics isomorphic structures can still lead to
  distinct predictions, e.g.\ sometimes the specific choice of a basis is
  significant.)
\itE The decision to take \calgo\ as the gauge group somehow gives undue
  prominence to the
  `constant gauge \trafo s', which do not possess a natural geometric meaning
  (recall that there is no canonical embedding of $G$ into \calg).
\itE The singular connection $A=0$ is the point about which
  one expands in ordinary perturbation theory.
  The restriction to the main stratum of \calm\ may therefore
  be incompatible with naive perturbation theory.
  \end{itemize}

\noindent Concerning the relevance of these choices,
first note that from a mathematical point of view one may choose one's
favourite structure and enjoy whatever results one is led to.
There remains, however, the question
whether there is any `physical' relevance
in the sense that -- at least in principle -- there are definite implications
for experimental predictions.
In order to answer this question, it is mandatory to have
insight into the theory at the quantum level.
(It is not possible to `turn on' and `switch off' quantum
corrections; this is an abuse of language which is unfortunately quite common,
and sometimes has disastrous effects).
As the understanding of many aspects of quantum \ym theory
so far is restricted to the realm of perturbation theory, again I can
present only a few isolated observations:
  \begin{itemize}
\itE The treatment of singular points in the \uline{quantum} theory
  potentially introduces further parameters, analogous to the construction
  of self-adjoint extensions of an operator on a Hilbert space that is not
  essentially self-adjoint.
\itE Various concepts in the quantum theory involve a description in
  terms of fiber bundles over the \confisp\ \calm. Examples are the
  determinant line bundle for chiral fermions that are coupled to the gauge
  fields \cite{algi,free3} (if this bundle does not have global sections,
  the theory has a
  chiral anomaly), and wave functionals in the Schr\"odinger picture.
  The very notion of fiber bundles and the like assumes, however, that
  \calm\ is a manifold.
\itE As already mentioned, the standard perturbation theory is an expansion
  about $A=0$. The orbit of $A=0$ is precisely the most singular point of
  \calm.
\itE In 2+1\,-dimensional \CS theory, reducible connections yield
  non-zero contributions to the Jones polynomial and to the Witten invariant
  of three-manifolds \cite{roza1+2}.
\itE In a combination of \ym and \CS theory in 2+1 dimensions, reducible
  connections give rise to non-trivial boundary conditions
  of the Schr\"odinger picture wave functionals \cite{afll2}.
  \end{itemize}

\sect{Outlook}

Let me now assume again that the description of \ym theories that I gave is
appropriate, and hence in particular that \calm\ has the stratified nature
described above. Some of the issues to be addressed in the future are then
as follows.
  \begin{itemize}
\itE An immediate task at the classical level is to classify the
  strata for structure groups $G$ other than SU(2), at least for SU(3).
  Ideally one would like to identify some `invariant', i.e.\ a machine
  which for any given connection $A$ tells to which stratum it belongs.
  Because of the presence of stabilizers which are isomorphic to
  non-abelian proper subgroups of $G$, this is quite difficult.
\itE Include matter fields. The singularities of the full configuration space
  then include those of the gauge fields tensored with the zero
  matter \confi. But some of these \confi s may have to be excluded because
  they would have infinite action (e.g.\ in the case of scalar matter fields
with a
  Higgs potential), while on the other hand there could exist further
singularities.
\itE Clarify the relation of the `cone over cones' structure of \calm\
  with a similar singularity structure that arises in a hamiltonian
formulation.
  The constraint subset of the pre-phase space is a submanifold except in
  the neighbourhood of reducible connections; after cutting out the singular
  points, the intersection of this constraint submanifold with the gauge
  slice is a symplectic submanifold of the pre-phase space, while including
  the reducible connections gives rise to a
  cone over cones structure \cite{monc2,arms,armm,gota}.
\ite Work out the relevant algebraic geometry aspects of stratified
  varieties. In a \findim\ context, similar aspects have been
  addressed e.g.\ in \cite{feRr2,brfe}.
\ite Desingularize \calm, e.g.\ by replacing the naive quotient $\cala/\calg$
  by the homotopy quotient associated to the classifying space $B\calg$
(compare
  \cite{bott4}), or by mimicking similar proposals for the space of
  Riemannian geometries \cite{fisc}.
\ite `Blow up' the singularities in a manner analogous to resolving
  orbifold singularities in complex \cite{GRha} and symplectic
\cite{AUdi,garo2}
  geometry. (This is one procedure by which one might introduce parameters.
  But situations are known, such as singularities of hypersurfaces in
  weighted projective spaces which appear in the description of string theory
  vacua, where the resolution is more or less canonical as far as
  the interesting `physical' quantities are concerned.) However,
  in real geometry, this can presumably only be achieved when the singularities
  can be described as the vanishing locus of some algebraic
  equation, which is not available here.
\ite Investigate the possibility that the proper quantum theory may smooth
  out the singularities automatically.
  As far fetched as this may sound, according to the results of \cite{seib.}
  such a phenomenon does occur for the moduli space of vacua
  of some supersymmetric gauge theories.
\ite Consider toy models, in particular models with \findim\ \confisp.
  In this context it may help to realize that
  in mathematics often much progress
  made by making use of algebraic rather than geometric tools (such as
  describing a manifold in terms of the algebra of smooth functions on it).
  A model which can be addressed in this spirit is \cs theory for which e.g.\
  the algebra of observables has been described explicitly in
  \cite{alsc}.
\ite Develop a measure theory on \calm\ so as to define the quantum
  theory by means of path integrals. this rather difficult,
  In the case of \calmo, a rigorous formulation of path integral measures has
  been achieved recently in \cite{baez3,asmm,asle3}
  (for earlier attempts, see e.g.\ \cite{bavi0,asmi}).
\end{itemize}
Finally, it is worth mentioning that in view of the complexity of
the structures described in this review the remarkable success of
naive perturbation theory for many aspects of \YM theory is quite mysterious.

\vskip9mm \small\noindent
{\bf Acknowledgement.} \\ Various results described in this paper have
been obtained in collaboration with M.G.\ Schmidt and C.\ Schweigert
\cite{fuss}. I am grateful to M.\ Asorey, K.\ Blaub\"ar, D.\ Giulini, and
D.\ Marolf for interesting discussions, and to C.\ Schweigert for carefully
reading the manu{\tiny?}script.
 \version\versionno \end{document}